\documentclass[aps,prl,a4paper,twocolumn,showpacs,floatfix,citeautoscript]{revtex4-1}
\usepackage{graphicx}
\usepackage{amsfonts}
\usepackage{amsmath}
\usepackage{amssymb}
\usepackage{bm}
\usepackage{dcolumn}
\usepackage{hyperref}
\usepackage[usenames]{color}
\usepackage{colortbl}
\definecolor{grey}{rgb}{0.8, 0.8, 0.8}
\makeatletter

    \def\CT@@do@color{%
      \global\let\CT@do@color\relax
            \@tempdima\wd\z@
            \advance\@tempdima\@tempdimb
            \advance\@tempdima\@tempdimc
    \advance\@tempdimb\tabcolsep
    \advance\@tempdimc\tabcolsep
    \advance\@tempdima2\tabcolsep
            \kern-\@tempdimb
            \leaders\vrule
                    \hskip\@tempdima\@plus  1fill
            \kern-\@tempdimc
            \hskip-\wd\z@ \@plus -1fill }
    \makeatother

\newcommand{\out}[1]{}

\hypersetup{pdfborder=0 0 0,colorlinks=true,citecolor=blue,linkcolor=blue}
\begin{document}

\title{A route to room temperature ferromagnetic ultrathin SrRuO$_3$ films}
\author{Liang Si, Zhicheng Zhong, Jan M. Tomczak and Karsten Held}
\affiliation{Institute of Solid State Physics, Vienna University of Technology, A-1040 Vienna, Austria}

\begin{abstract} 
Experimental efforts to stabilize ferromagnetism in ultrathin films of transition metal oxides have so far failed, despite
expectations based on density functional theory (DFT) and DFT+U.
Here, we investigate one of the most promising materials, SrRuO$_3$, and
include  correlation effects beyond DFT by means of  dynamical mean field theory. In agreement with experiment we find
an intrinsic thickness limitation for metallic ferromagnetism in SrRuO$_3$ thin films. 
Indeed, we demonstrate that the realization of ultrathin ferromagnetic films is out of reach of standard thin-film techniques.
Proposing charge carrier doping as a new route to manipulate thin films, we predict room temperature ferromagnetism
in electron-doped SrRuO$_3$ ultra thin films.
\end{abstract}
\maketitle


Thin films and heterostructures of the 4$d$ perovskite SrRuO$_3$ (SRO) are intensively studied and used,
in particular, as gate electrodes for novel oxide-based electronic devices \cite{junquera2003critical, stengel2006origin}.
The reason for this is that SRO  is a conductor with good thermal properties \cite{lee2004thermal} and high chemical stability, allowing for epitaxial growth on various substrates, as well as the combining
with other perovskite-based materials to form complex heterostructures \cite{eom1992single, koster2012structure}. 
In the bulk,
SRO is a ferromagnetic (FM) metal with a, for a 4$d$ oxide, remarkably high 
Curie temperature,  $T_C =160K$, and an experimental magnetic moment in the range of 0.8 to 1.6 $\mu_B$ \cite{PhysRevB.54.R756, PhysRevB.56.321, allen1996transport, felner2006mossbauer}.
SRO further  attracted  fundamental research interests regarding, among others,  magnetic monopoles \cite{fang2003anomalous}, non-Fermi Liquid \cite{laad2001origin}, spin freezing \cite{werner2008spin}, and the debate of itinerant \cite{jeong2013temperature} versus localized magnetism \cite{shai2013quasiparticle}.

However, the FM moment and Curie temperature get dramatically suppressed below a sample thickness
of $4$ unit cells \cite{toyota2005thickness,toyota2006ferromagnetism,xia2009critical,chang2009fundamental}, and 
eventually single unit cell SRO films turn antiferromagnetic (AF) and insulating \cite{xia2009critical,chang2009fundamental}.
This led to the pertinent question whether there is a fundamental thickness limit for ferromagnetism \cite{chang2009fundamental},
and concerted efforts to stabilize ferromagnetism in ultrathin SRO films by compressive and tensile strain or capping layers \cite{verissimo2012highly,gupta2014strain}. 
However, hitherto ferromagnetism in single unit cell films remains unattainable for SRO or any other oxide material, even in a heterostructured setup. 

On the theoretical side, previous attempts to understand the electronic structure and the transition to an AF insulator resorted to density functional theory (DFT) and the static mean-field DFT+U approach.
The former failed to reproduce the transition \cite{rondinelli2008electronic}, while the latter found a transition to an AF insulating state below four layers when assuming an artificial RuO$_2$ terminated surface \cite{mahadevan2009evolution}, while experimentally samples are found to have a SrO termination \cite{koster2012structure}. 
DFT+U further predicted a spin-polarized highly confined half-metallic state for an SRO mono-layer when either sandwiched with SrTiO$_3$ (STO) \cite{verissimo2012highly} or grown on a strained STO substrate \cite{gupta2014strain}. However, such a state could not be confirmed in experiment \cite{bern2013structural}. The apparent discrepancy between experiments and results from standard band-structure methods calls for a more sophisticated treatment of electronic correlation effects. 
Indeed already in the bulk, SRO displays signatures of electronic correlations, such as many-body satellites in photoemission or the violation of the Ioffe-Regel limit in the resistivity \cite{allen1996transport, emery1995superconductivity}.
Hence, SRO is to be considered an --at least-- moderately correlated system. Note that a dimensional reduction/geometric constraints in thin films can be expected to further enhance electronic correlations.

For a better and unbiased treatment of these correlations effects in various SRO films and heterostructure setups,
we employ realistic DFT + dynamical mean-field theory (DMFT) \cite{georges1996dynamical, kotliar2004strongly, held2007electronic, anisimov1997first, kotliar2006electronic} calculations.
Our main findings are:
(1) Both the SRO mono-layer and bi-layer are AF insulators.
(2) We demonstrate that standard thin film manipulation techniques such as strain and surface capping can neither restore ferromagnetism nor metallicity to a  SRO mono-layer; interestingly, we find that surface capping pushes the AF insulator towards a paramagnetic (PM) insulator.
(3) With new insight regarding the microscopic origin for the transition, we identify carrier doping as the best option to generate FM properties that are on a par with those of the bulk. We find the FM moments 
of doped SRO films to be stable even at room temperature, heralding a great potential for technological applications.

\begin{figure}
\includegraphics[width=2.7in]{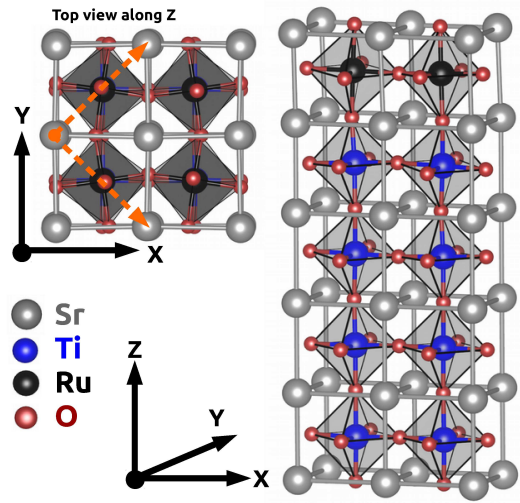}
\caption{Right: Structure of a SRO mono-layer grown on 4 layers of STO. Upper left: Top view of the same structure. The indicated $\sqrt2\times\sqrt2$ 
supercell was adopted for allowing AF-ordering 
in each RuO$_2$ layer. Lower left: atomic labels and coordinate system (Figures drawn with the Vesta code \cite{momma2011vesta}).} 
\label{Fig1}
\end{figure}

{\em Method.} 
We use the experimental orthorhombic crystal structure of SRO \cite{jones1989structure} for the various setups of bulk, films and heterostructures. In the films and heterostructure both the internal positions and lattice constant are relaxed; the in-plane lattice constants in films are fixed to the experimental ones of STO. Fig.~\ref{Fig1} exemplary shows the SRO mono-layer grown on 4 layers of STO substrate. The atomic relaxations are carried out with the VASP program package \cite{PhysRevB.48.13115, Kresse199615} using the PBE functional \cite{perdew1996generalized}. For the optimized atomic positions, we subsequently perform WIEN2K \cite{blaha2001wien2k} electronic structure calculations with the mBJ exchange \cite{tran2009accurate} and PBE correlation functional \footnote{As detailed in the Supplemental Material, the exchange included in mBJ does not notably affect the $t_{2g}$ bandwidth, but it improves on the inter-orbital separation for states not included in the DMFT. It is in this sense a poor man's version of QSGW+DMFT \cite{tomczak2014qsgw+}.}, and a Wannier function projection onto maximally localized  \cite{marzari2012maximally} $t_{2g}$ Wannier orbitals \cite{mostofi2008wannier90} using the  Wien2Wannier program package \cite{kunevs2010wien2wannier}. This $t_{2g}$  Hamiltonian is supplemented by a local Kanamori interaction and solved within DMFT using  Wien2Dynamics \cite{parragh2012conserved}, employing a 
hybridization expansion \cite{gull2011continuous} continuous-time quantum Monte Carlo (CTQMC) algorithm. 
For the Coulomb interaction strengths, we adopt a Hund's exchange  ($J=0.3\,$eV), intra- ($U=3.0\,$eV) and inter-orbital  Coulomb repulsion ($U'=2.4\,$eV). These values are  chosen not only because they are in between the constrained random phase approximation (cRPA) values for (i) free standing cubic SRO mono-layer (0.3eV, 3.5eV,  and 2.9eV) and (ii)  orthorhombic bulk
and (0.3eV, 2.3eV, and 1.7eV), but also because they reproduce the FM metallic state for orthorhombic bulk SRO.

{\em Bulk SrRuO$_3$.} 
The moderately correlated electronic structure of {\it bulk} SRO was successfully captured in both many-body perturbation theory \cite{hadipour2011electron} and realistic DMFT calculations \cite{jakobi2011lda+,graanas2014electronic}. Also DFT calculations correctly predict that SRO is an itinerant ferromagnet with a moment ranging from 1.5 to 1.6$\mu_B$ \cite{allen1996transport,singh1996electronic}; similar moments have also been obtained within  DFT+DMFT \cite{graanas2014electronic,kim2015nature}. 
Using DFT+DMFT, we indeed find orthorhombic bulk SRO to be a FM metal with 
orbital occupations of 0.867 (0.466) for the majority (minority) spin of all three $t_{2g}$  orbitals at $T=100$K. This corresponds to a FM magnetic moment of $\sim 1.2\mu_B$. A GdFeO$_3$-type distortion in which the corner-sharing octahedral tilt around the $y$-axis and rotate around the $z$-axis lifts, in principle, the $t_{2g}$ degeneracy. The effect on the crystal field and orbital occupations of orthorhombic SRO is however minute. Our DFT+DMFT finds SRO to be a PM metal above $T_c\sim$ 150K which is close to the experimental Curie temperature of 160K (cf. Fig.~\ref{Fig3} below).

\begin{figure}
\includegraphics[width=\columnwidth]{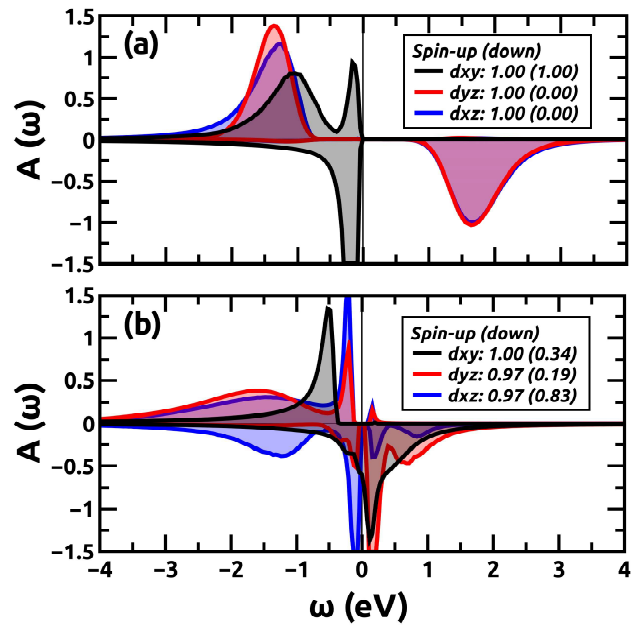}
\caption{DFT+DMFT spectral functions of (a) the AF SRO mono-layer on a STO substrate and
(b) the  FM  doped (4.3 electrons/Ru) superlattice  (STO)$_5$:(SRO)$_1$ at 150K. Insets: electronic occupations.}
\label{Fig2}
\end{figure}

{\em Thin films.}
We now consider SRO grown on STO, and study the evolution of the electronic structure when reducing the number of SRO layers.  We find that FM is suppressed: the mono- and bi-layer SRO on  STO are AF insulators,
in congruence with experiments \cite{bern2013structural,chang2009fundamental} that show a dramatic drop in the FM moment and an insulating behavior for 4 layers or lower \cite{toyota2005thickness,toyota2006ferromagnetism,xia2009critical,bern2013structural,chang2009fundamental}. 
Indicative of an itinerant origin of ferromagnetism, the critical thickness of the magnetic and electronic transition coincide.
Fig.~\ref{Fig2} (a) shows the spectral function of the mono-layer. The system is gapped by $\sim$1.0eV  and displays a large orbital polarization: the $xy$ orbital is fully filled, and the $xy$ and $xz$ orbitals are half-filled and fully spin-polarized,  resulting in an AF moment of $\sim2\mu_B$. This finding is supported by recent exchange bias measurements \cite{xia2009critical}. 

We note that for the particular case of the SRO mono-layer, also LDA+U \cite{gupta2014strain} can seemingly give a qualitative correct picture, as the system is orbitally and spin-polarized.
However, the underlying physics is very different: When heating the mono-layer above its N{\'e}el temperature within DMFT, the system remains insulating at non-integer filling 
[0.88 (0.88), 0.56 (0.56), and 0.56 (0.56) for the spin up (down) $xy$, $yz$ and $xz$ orbitals at 1000K]. 
This complex Mott physics \cite{PhysRevB.78.045115} reveals that the AF insulating phase is beyond a simple Slater description, and thus not describable by LDA+U.

{\em Physical origin of transition}. Let us now investigate the microscopic origin of the FM-metal to AF-insulator transition.
Whereas the crystal field splitting is minute for the bulk, in the SRO mono-layer the $xy$-orbital is energetically lower than the $yz$ and $xz$ orbitals, because the (cubic) crystal symmetry is strongly broken at the surface. This is already the case for the DFT Wannier Hamiltonian, but correlation effects boost the crystal field splitting \cite{PhysRevB.70.205116,PhysRevB.76.085127,PhysRevB.78.045115} of the SRO mono-layer (see Supplemental materials Table \ref{table1}). Therewith, the $xy$-orbital become essentially fully occupied, and the two remaining electrons occupy the $yz$ and $xz$ orbitals: The single-layer SRO is an effective half-filled two-band system, favourable to AF order. 

{\em Tuning the properties of the SRO mono-layer.} 
The prime motivation for SRO-based thin films are the advantageous properties of the FM metallic {\it bulk}. However, the desired features such as the magnetic moment strongly decrease for thinner films and eventually ultrathin films become non-FM, in agreement with our calculations. A natural question is whether the bulk properties can be restored, at least partially, by tuning the geometry of the films.

First we discuss the influence of straining/tensioning the mono-layer. This can be realized experimentally by choosing an appropriate substrate. Indeed, previous DFT+U calculations \cite{gupta2014strain} predicted a strain-induced FM half-metallic state for the SRO mono-layer. DFT+DMFT, however, does not show any tendency towards a FM half-metallic state at least for   realistic $U$ values,   see Table II in the Supplemental Materials. Also, the effective crystal-field splitting $\Delta_{eff}$, shown in Table \ref{table1} of Supplemental Materials, can only be tuned slightly through straining/tensioning. 

\begin{table}
\caption{Magnetic and conductive states of bulk SRO, SRO mono-layer (also under 0.5\% compression and 0.5\% tension),
SRO bi-layers and (STO)$_5$:(SRO)$_1$ superlattice (capped mono-layer), as obtained from spin-polarized DFT and DFT+DMFT calculations in comparison with experiment.
FM-M: ferromagnetic metal,  AF-I: antiferromagnetic insulator; NM-I: non-ferromagnetic insulator (the magnetic nature of the experimental non-ferromagnetic state has not been fully determined yet; the exchange bias behavior hints at antiferromagnetism \cite{xia2009critical}).}
\begin{tabular}{ l|l|l|l }
\hline
\hline
System & DFT & DMFT & EXP \\
\hline
bulk & FM-M & FM-M  & FM-M \cite{PhysRevB.54.R756}\\
mono-layer 100.5\% &FM-M & AF-I &  - \\
mono-layer 100\% & FM-M & AF-I  & NM-I \cite{xia2009critical}\\
mono-layer 99.5\% & FM-M & AF-I & - \\
bi-layer & FM-M & AF-I & NM-I \cite{toyota2005thickness,toyota2006ferromagnetism} \\
superlattice & FM-M  & AF-I & NM-I \cite{bern2013structural}\\
hole doped superlattice & FM-M  & FM-M & - \\
electrons doped superlattice & FM-M  & FM-M & - \\
\hline
\hline
\end{tabular}
\label{table1}
\end{table}

Another way to influence the crystal-field splitting is through the deployment of capping layers.
Here, we study the effect of capping the SRO mono-layer with additional layers of STO. Specifically, we consider a (STO)$_5$:(SRO)$_1$ superlattice \cite{verissimo2012highly} consisting of 5 layers of STO alternating with a mono-layer of SRO. This restores, at least partially, the hopping amplitudes in the out-of-plane direction. Compared to the SRO mono-layer, the DFT crystal field splitting, shown in Supplemental materials Table \ref{table1}, is now much smaller (-0.05eV) approaching the negligible value of the bulk. As a result, the $t_{2g}$ orbital occupations are more balanced. However, this causes only a slight reduction of the AF magnetic moment (1.92$\mu_B$ at 150K and 1.48$\mu_B$ at 300K, cf. Fig.~\ref{Fig3} (b)) with respect to the un-capped mono-layer.
Our finding of a non-FM insulating state with a gap of $\sim$1.0eV for the capped mono-layer is consistent with experiments \cite{tian2007epitaxial, bern2013structural}, where it was concluded that SRO capped by STO leads to a insulator without a net moment. We note that previous DFT+U calculations \cite{verissimo2012highly} instead predicted a FM half-metal, at variance with experiment.

The above calculations reveal that the non-FM insulating state is a robust feature of the SRO mono-layer. 
Only for an unrealistically small $U$-to-bandwidth ratio a FM phase can be stabilized (see Supplemental materials). 
In consequence, none of the standard manipulation strategies available to the production of thin films can tune this ratio sufficiently to induce ferromagnetism to the SRO mono-layer. This stability explains why all experimental efforts to create ultrathin ferromagnetic films have so far been unsuccessful.

\begin{figure}
\includegraphics[width=\columnwidth]{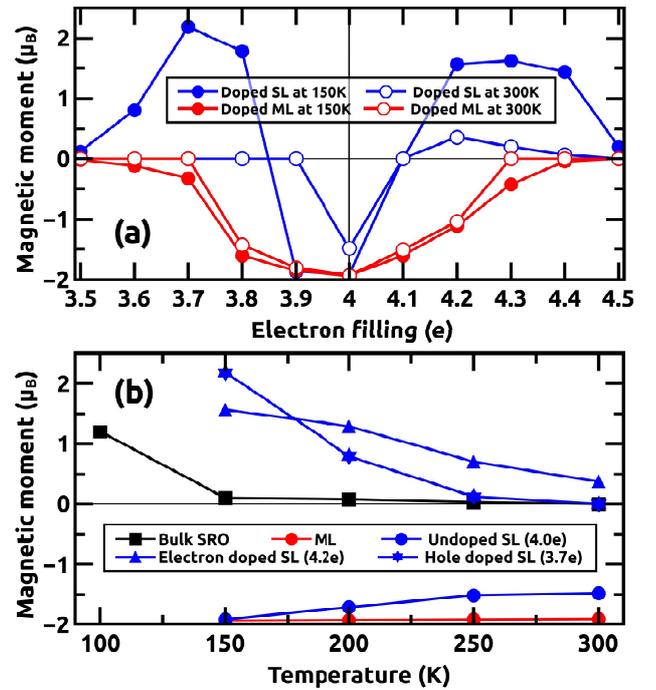}
\caption{(a) Magnetic moment of SRO mono-layer (ML) and (STO)$_5$:(SRO)$_1$ superlattice (SL) vs.\ doping at 150K and 300K (positive [negative] moments denote FM [AF] ordering). (b) Magnetic moments vs. temperature for orthorhombic bulk SRO, mono-layer, undoped and doped (STO)$_5$:(SRO)$_1$ superlattice.}
\label{Fig3}
\end{figure}

{\em Doping.} Here we propose an alternative route to achieve ultrathin FM films: doping. 
This strategy may seem counter-intuitive at first glance, as carrier doping causes a deterioration of the desired properties in the {\it bulk}. As we shall see, the situation for ultra thin films is different:
Using the virtual crystal approximation to simulate carrier doping within DFT+DMFT, we obtain for the SRO mono-layer and the (STO)$_5$:(SRO)$_1$ superlattice the magnetic moments shown in Fig.~\ref{Fig3}. For the SRO mono-layer a significant doping corresponding to 3.5 and 4.5 electrons/site is needed to turn the AF state into a PM at both low (150K) and high temperature (300K), see Fig.~\ref{Fig3} (a). The AF magnetic moment is essentially symmetric around the filling with four electrons. The reason for this is the previously mentioned crystal field effect that results in an almost fully occupied $xy$ orbital and a half-filled (particle-hole symmetric) $yz/xz$ orbital doublet. We note that in all cases, the doping away from four electrons induces a metallic state.

Let us now turn to the more important case, the (STO)$_5$:(SRO)$_1$ superlattice:
At low temperatures, e.g.\ at 150K as shown in  Fig.~\ref{Fig3} (a), both hole and electron doping can induce strong FM states. In the case of electron doping (filling $>4$) the FM state is accompanied by an alternating orbital ordering of the $xz$/$yz$ minority spin. The spectral function corresponding to 4.3 electrons/site is shown in Fig.~\ref{Fig2} (b). 
There, one Ru site has the orbital occupations for up (down) spin: $xy$ 1.00 (0.34), $yz$ 0.97 (0.83), $xz$ 0.97 (0.19); while for the second Ru site: $xy$ 1.00 (0.34), $yz$ 0.97 (0.19), $xz$ 0.97 (0.83). For hole doping, e.g. at 3.7 electrons, our DMFT results indicate that the system is a FM half-metal with a moment of 2.20 $\mu_B$/Ru at 150K, see Fig.~\ref{Fig3} (b) and the Supplemental Materials for the corresponding spectral functions. The half-metallic behavior makes this setup a prospective candidate for spintronics applications.
To put this finding into perspective, we recall that bulk SRO has an FM moment of 2$\mu_B$/Ru or less, and an experimental (theoretical) Curie temperature of ``only'' 160K (150K).
One might thus wonder whether the ferromagnetism of the doped supercell is actually superior to the hailed properties of stoichiometric bulk SRO, which we were striving to restore.
To investigate this, we perform calculations as a function of temperature, see Fig.~\ref{Fig3} (b).
We find that for both hole and electron doping, magnetic moments and Curie temperatures  of the supercell are remarkably higher than for bulk SRO. 
In particular, for 4.2 electrons/Ru site, a sizable magnetic moments survives up to room temperature 300K, see Fig.~\ref{Fig3} (b). The magnetization curve has a similar shape as for the bulk \cite{kim2015nature}, despite the much higher $T_c$ and the orbital-ordering.
In the Supplemental Material we also go beyond the virtual crystal approximation and show that a (STO)$_5$:(La$_{0.25}$Sr$_{0.75}$RuO$_3$)$_1$ superlattice is indeed FM. 
Our findings pave the road for realizing FM oxide devices that can be operated at room temperature.


{\em Conclusion.} Including many-body effects by means of DFT+DMFT, we show that the SRO mono-layer is an AF Mott insulator owing to a correlation enhanced crystal-field splitting at the surface/interface. 
While the bare (one-particle) crystal-field splitting can be tuned to almost zero by STO capping layers, electronic correlations are still strong enough to boost the orbital separation so that also the capped SRO layer 
remains an AF insulator. A FM metallic state is only realized for an interaction-to-bandwidth ratio that cannot be realized by experiment. This explains why no ultrathin FM films could be stabilized in experiment to date. Given the robustness of the AF state of SRO mono-layer setups to standard thin film manipulation techniques, we propose an alternative route to realize a FM state:
Our study suggests that carrier doping drives ultrathin SRO films capped with STO into a strong FM state, whose ordered moment and Curie temperature even exceed the values realized in stoichiometric bulk SRO.
To achieve the long-standing quest for a FM ultrathin film in practice, we consider inducing oxygen or Sr vacancies \cite{kim2005situ} or doping potassium into STO:SRO superlattices \cite{bern2013structural} as the most promising means.

Our study also opens a new, general, perspective: Producing heterostructures based on materials with optimized {\it bulk} properties (e.g.\ stoichiometric SRO) is actually not always the optimal way for achieving those properties in a {\it film} geometry. Indeed the electronic structure of the thin film is so different from the bulk that it can be viewed as a completely different material. A manipulation (in our case doping) that decreases the quality of the bulk, can in fact enhance the sought-after property (FM magnetic moment) for the film setup. This suggests in turn that rather inconspicuous bulk materials 
might actually be good candidates for specific functionalities when deployed in a film or heterostructure. With this observation the repertoire of materials to be evaluated for oxide-electronics applications
is significantly enlarged.

{\em Acknowledgments.} LS, JMT and KH acknowledge financial support by European Research Council under the European Union's Seventh Framework Program (FP/2007-2013)/ERC through grant agreement n.\ 306447, ZZ acknowledges financial support by the Austrian Science Fund through the SFB ViCoM F4103-N13. LS also thanks the support from Doctoral School Solids4Fun (Building Solids for Function). Calculations have been done on the Vienna Scientific Cluster~(VSC).

%

\end{document}